\batchmode
\makeatletter
\def\input@path{{"/home/jacob/Documents/Work/My Papers/Stochastic Processes and Quantum Theory (2023)/"}}
\makeatother
\documentclass[twoside,twocolumn,english,prl,superscriptaddress,nobibnotes,nofootinbib]{revtex4-2}
\usepackage[latin9]{inputenc}
\usepackage{babel}
\usepackage{mathtools}
\usepackage{amsmath}
\usepackage{amssymb}
\usepackage{graphicx}
\usepackage[pdftex,unicode=true,
 bookmarks=true,bookmarksnumbered=true,bookmarksopen=false,
 breaklinks=false,pdfborder={0 0 0},pdfborderstyle={},backref=false,colorlinks=false]
 {hyperref}
\hypersetup{
 pdfauthor={Jacob A. Barandes},
 pdfsubject={Physics},
 pdfkeywords={physics}}

\makeatletter

\let\originalleft\left
\let\originalright\right
\renewcommand{\left}{\mathopen{}\mathclose\bgroup\originalleft}
\renewcommand{\right}{\aftergroup\egroup\originalright}

\def\smalloverbrace#1{\mathop{\vbox{\m@th\ialign{##\crcr%
      \noalign{\kern3\p@}%
      \tiny\downbracefill\crcr\noalign{\kern3\p@\nointerlineskip}%
      $\hfil\displaystyle{#1}\hfil$\crcr}}}\limits}

\def\smallunderbrace#1{\mathop{\vtop{\m@th\ialign{##\crcr
   $\hfil\displaystyle{#1}\hfil$\crcr
   \noalign{\kern3\p@\nointerlineskip}%
   \tiny\upbracefill\crcr\noalign{\kern3\p@}}}}\limits}

\DeclareMathAlphabet{\mymathbb}{U}{bbold}{m}{n}

\makeatother

\begin{document}
\title{New Prospects for a Causally Local Formulation of Quantum Theory}
\author{Jacob A. Barandes}
\email{jacob\_barandes@harvard.edu}

\affiliation{Jefferson Physical Laboratory, Harvard University, Cambridge, MA 02138}
\affiliation{Department of Philosophy, Harvard University, Cambridge, MA 02138}
\date{\today}
\begin{abstract}
It is difficult to extract reliable criteria for causal locality
from the limited ingredients  found in textbook quantum theory.
In the end, Bell humbly warned that his eponymous theorem was based
on criteria that ``should be viewed with the utmost suspicion.''
Remarkably, by stepping outside the wave-function paradigm, one can
reformulate quantum theory in terms of old-fashioned configuration
spaces together with \textquoteleft unistochastic\textquoteright{}
 laws. These unistochastic laws take the form of directed conditional
probabilities, which turn out to provide a hospitable foundation for
encoding microphysical causal relationships. This unistochastic reformulation
provides quantum theory with a simpler and more transparent axiomatic
 foundation, plausibly resolves the measurement problem, and deflates
various exotic claims about superposition, interference, and entanglement.
Making use of this reformulation, this paper introduces a new principle
of causal locality that is intended to improve on Bell's criteria,
and shows directly that systems that remain at spacelike separation
cannot exert causal influences on each other, according to that new
principle. These results therefore lead to a general hidden-variables
interpretation of quantum theory that is arguably compatible with
causal locality.
\end{abstract}
\maketitle

\noindent \begin{center}
\global\long\def\quote#1{``#1"}%
\global\long\def\apostrophe{\textrm{'}}%
\global\long\def\slot{\phantom{x}}%
\global\long\def\eval#1{\left.#1\right\vert }%
\global\long\def\keyeq#1{\boxed{#1}}%
\global\long\def\importanteq#1{\boxed{\boxed{#1}}}%
\global\long\def\given{\vert}%
\global\long\def\mapping#1#2#3{#1:#2\to#3}%
\global\long\def\composition{\circ}%
\global\long\def\set#1{\left\{  #1\right\}  }%
\global\long\def\setindexed#1#2{\left\{  #1\right\}  _{#2}}%

\global\long\def\setbuild#1#2{\left\{  \left.\!#1\,\right|\,#2\right\}  }%
\global\long\def\complem{\mathrm{c}}%

\global\long\def\union{\cup}%
\global\long\def\intersection{\cap}%
\global\long\def\cartesianprod{\times}%
\global\long\def\disjointunion{\sqcup}%

\global\long\def\isomorphic{\cong}%

\global\long\def\setsize#1{\left|#1\right|}%
\global\long\def\defeq{\equiv}%
\global\long\def\conj{\ast}%
\global\long\def\overconj#1{\overline{#1}}%
\global\long\def\re{\mathrm{Re\,}}%
\global\long\def\im{\mathrm{Im\,}}%

\global\long\def\transp{\mathrm{T}}%
\global\long\def\tr{\mathrm{tr}}%
\global\long\def\adj{\dagger}%
\global\long\def\diag#1{\mathrm{diag}\left(#1\right)}%
\global\long\def\dotprod{\cdot}%
\global\long\def\crossprod{\times}%
\global\long\def\Probability#1{\mathrm{Prob}\left(#1\right)}%
\global\long\def\Amplitude#1{\mathrm{Amp}\left(#1\right)}%
\global\long\def\cov{\mathrm{cov}}%
\global\long\def\corr{\mathrm{corr}}%

\global\long\def\absval#1{\left\vert #1\right\vert }%
\global\long\def\expectval#1{\left\langle #1\right\rangle }%
\global\long\def\op#1{\hat{#1}}%

\global\long\def\bra#1{\left\langle #1\right|}%
\global\long\def\ket#1{\left|#1\right\rangle }%
\global\long\def\braket#1#2{\left\langle \left.\!#1\right|#2\right\rangle }%

\global\long\def\parens#1{(#1)}%
\global\long\def\bigparens#1{\big(#1\big)}%
\global\long\def\Bigparens#1{\Big(#1\Big)}%
\global\long\def\biggparens#1{\bigg(#1\bigg)}%
\global\long\def\Biggparens#1{\Bigg(#1\Bigg)}%
\global\long\def\bracks#1{[#1]}%
\global\long\def\bigbracks#1{\big[#1\big]}%
\global\long\def\Bigbracks#1{\Big[#1\Big]}%
\global\long\def\biggbracks#1{\bigg[#1\bigg]}%
\global\long\def\Biggbracks#1{\Bigg[#1\Bigg]}%
\global\long\def\curlies#1{\{#1\}}%
\global\long\def\bigcurlies#1{\big\{#1\big\}}%
\global\long\def\Bigcurlies#1{\Big\{#1\Big\}}%
\global\long\def\biggcurlies#1{\bigg\{#1\bigg\}}%
\global\long\def\Biggcurlies#1{\Bigg\{#1\Bigg\}}%
\global\long\def\verts#1{\vert#1\vert}%
\global\long\def\bigverts#1{\big\vert#1\big\vert}%
\global\long\def\Bigverts#1{\Big\vert#1\Big\vert}%
\global\long\def\biggverts#1{\bigg\vert#1\bigg\vert}%
\global\long\def\Biggverts#1{\Bigg\vert#1\Bigg\vert}%
\global\long\def\Verts#1{\Vert#1\Vert}%
\global\long\def\bigVerts#1{\big\Vert#1\big\Vert}%
\global\long\def\BigVerts#1{\Big\Vert#1\Big\Vert}%
\global\long\def\biggVerts#1{\bigg\Vert#1\bigg\Vert}%
\global\long\def\BiggVerts#1{\Bigg\Vert#1\Bigg\Vert}%
\global\long\def\ket#1{\vert#1\rangle}%
\global\long\def\bigket#1{\big\vert#1\big\rangle}%
\global\long\def\Bigket#1{\Big\vert#1\Big\rangle}%
\global\long\def\biggket#1{\bigg\vert#1\bigg\rangle}%
\global\long\def\Biggket#1{\Bigg\vert#1\Bigg\rangle}%
\global\long\def\bra#1{\langle#1\vert}%
\global\long\def\bigbra#1{\big\langle#1\big\vert}%
\global\long\def\Bigbra#1{\Big\langle#1\Big\vert}%
\global\long\def\biggbra#1{\bigg\langle#1\bigg\vert}%
\global\long\def\Biggbra#1{\Bigg\langle#1\Bigg\vert}%
\global\long\def\braket#1#2{\langle#1\vert#2\rangle}%
\global\long\def\bigbraket#1#2{\big\langle#1\big\vert#2\big\rangle}%
\global\long\def\Bigbraket#1#2{\Big\langle#1\Big\vert#2\Big\rangle}%
\global\long\def\biggbraket#1#2{\bigg\langle#1\bigg\vert#2\bigg\rangle}%
\global\long\def\Biggbraket#1#2{\Bigg\langle#1\Bigg\vert#2\Bigg\rangle}%
\global\long\def\angs#1{\langle#1\rangle}%
\global\long\def\bigangs#1{\big\langle#1\big\rangle}%
\global\long\def\Bigangs#1{\Big\langle#1\Big\rangle}%
\global\long\def\biggangs#1{\bigg\langle#1\bigg\rangle}%
\global\long\def\Biggangs#1{\Bigg\langle#1\Bigg\rangle}%

\global\long\def\vec#1{\mathbf{#1}}%
\global\long\def\vecgreek#1{\boldsymbol{#1}}%
\global\long\def\idmatrix{\mymathbb{1}}%
\global\long\def\projector{P}%
\global\long\def\permutationmatrix{\Sigma}%
\global\long\def\densitymatrix{\rho}%
\global\long\def\krausmatrix{K}%
\global\long\def\stochasticmatrix{\Gamma}%
\global\long\def\lindbladmatrix{L}%
\global\long\def\dynop{\Theta}%
\global\long\def\timeevop{U}%
\global\long\def\hadamardprod{\odot}%
\global\long\def\tensorprod{\otimes}%

\global\long\def\inprod#1#2{\left\langle #1,#2\right\rangle }%
\global\long\def\normket#1{\left\Vert #1\right\Vert }%
\global\long\def\hilbspace{\mathcal{H}}%
\global\long\def\samplespace{\Omega}%
\global\long\def\configspace{\mathcal{C}}%
\global\long\def\phasespace{\mathcal{P}}%
\global\long\def\spectrum{\sigma}%
\global\long\def\restrict#1#2{\left.#1\right\vert _{#2}}%
\global\long\def\from{\leftarrow}%
\global\long\def\statemap{\omega}%
\global\long\def\degangle#1{#1^{\circ}}%
\global\long\def\trivialvector{\tilde{v}}%
\global\long\def\eqsbrace#1{\left.#1\qquad\right\}  }%
\par\end{center}

\section{Introduction\label{sec:Introduction}}

In physics, \textquoteleft locality\textquoteright{} can refer to
any of several distinguishable concepts. What follows is a non-exhaustive
list of historically important examples.
\begin{itemize}
\item In physical theories like Newtonian mechanics that involve forces,
one can ask whether those forces are limited by the speed of light,
or instead consist of faster-than-light \emph{action at a distance}.
A well-known case of action at a distance is the Newtonian gravitational
force $F_{\textrm{g}}=Gm_{1}m_{2}/\verts{\vec r_{1}-\vec r_{2}}^{2}$
between two spherically symmetric bodies with respective masses $m_{1}$
and $m_{2}$, and with respective centers of mass located at positions
$\vec r_{1}$ and $\vec r_{2}$, where $G$ is Newton's constant.
The status of this form of nonlocality is somewhat murkier in textbook
formulations of quantum theory, in which forces do not appear to play
a fundamental role.
\item A physical theory is \emph{signal-local}~\citep{Skyrms:1982cdalc,Skyrms:1984elfm}
if it does not permit the transmission of controllable signals or
messages faster than light. In principle, there are no constraints
in Newtonian mechanics that would preclude sending superluminal signals\textemdash say,
by exploiting the action-at-a-distance features of Newtonian gravitational
forces. Newtonian mechanics is therefore presumably \emph{signal-nonlocal}.
By contrast, the aptly named \emph{no-communication theorem}~\citep{GhirardiRiminiWeber:1980agaastttqmmp,Jordan:1983qcdnts}
ensures that appropriately defined quantum systems\textemdash such
as local quantum fields\textemdash cannot be used to send superluminal
signals, so these quantum systems are signal-local.
\item The \emph{cluster decomposition principle}~\citep{WichmannCrichton:1963cdpotsm,Weinberg:1996tqtfi}
is the condition that correlation functions for a physical system
consisting of widely separated constituent subsystems should factorize
into a product of correlation functions for each of those individual
subsystems. This condition ensures that the statistical behavior of
nearby physical systems does not depend on the inaccessible details
of other systems that are very far away, assuming the absence of any
initial correlations between the nearby and faraway systems.
\item For a local quantum field theory, one typically imposes \emph{microcausality conditions}~\citep{Weinberg:1996tqtfi},
which require that bosonic field operators should commute at spacelike
separation, and that fermionic field operators should anticommute
at spacelike separation. Among other consequences, these microcausality
conditions ensure that local observables at spacelike separation are
capable of being statistically uncorrelated.
\item At the level of mereology, a spatially extended physical entity that
is fully reducible to spatially local parts is said to be \emph{separable},
and is otherwise said to be \emph{nonseparable} or \emph{holistic}~\citep{Howard:1985eolas,Howard:1989hsatmiotbe}.
\end{itemize}

This paper will be concerned with a different type of locality, called
\emph{causal locality}, which will be taken to consist of the following
statement: \begin{equation}
\left.\begin{minipage}{\columnwidth}
\leftskip=10pt 
\rightskip=60pt 

Causal influences should not be able to propagate faster than light.

\end{minipage}\hspace{-50pt}\right\}
\label{eq:DefCausalityLocality}
\end{equation}

Going back at least to the work of Albert Einstein, Boris Podolsky,
and Nathan Rosen in 1935~\citep{EinsteinPodolskyRosen:1935cqmdprbcc},
and continuing through the work of John Bell in the 1960s and beyond~\citep{Bell:1964oeprp,ClauserHorneShimonyHolt:1969pettlhvt,Bell:1975ttolb,Bell:1981bssatnor,GreenbergerHorneZeilinger:1989gbbt,Bell:1990lnc,Mermin:1990qmr},
there has been an ongoing debate over whether quantum theory is causally
local in this sense. A major challenge for all such arguments is that
causal locality expressly depends on the notion of a \textquoteleft causal
influence,\textquoteright{} which is a notoriously difficult concept
to define rigorously. One of the main goals of this paper will be
to address this difficulty directly, as a stepping stone toward arguing
that a specific new formulation of quantum theory~\citep{Barandes:2023tsqc,Barandes:2023tsqt}
is, in fact, causally local.

After a high-level overview of the Einstein-Podolsky-Rosen (EPR) argument
and Bell's subsequent work in Section~\ref{sec:Einstein-Podolsky-Rosen-and-Bell},
Section~\ref{sec:Bell's-Principle-of-Local-Causality} will continue
with a detailed analysis of Bell's results and their assumed criteria
for causal locality. Section~\ref{sec:The-Unistochastic-Formulation-of-Quantum-Theory}
will then review a new \emph{unistochastic formulation} of quantum
theory, based on \textquoteleft unistochastic\textquoteright{} microphysical
laws~\citep{Barandes:2023tsqc,Barandes:2023tsqt}. Section~\ref{sec:Bayesian-Networks-and-Causation}
will introduce salient topics related to causality from the theory
of Bayesian networks, and then, inspired in part by those ideas, Section~\ref{sec:A-Microphysical-Account-of-Causation}
will recast the unistochastic formulation in causal terms. Section~\ref{sec:An-Improved-Principle-of-Causal-Locality}
will show that this overall approach makes possible an improved criterion
for causal locality. Section~\ref{sec:Revisiting-the-Einstein-Podolsky-Rosen-Argument}
will then argue that the unistochastic formulation is causally local
according that improved criterion. Section~\ref{sec:Conclusion}
will conclude with a summary and a discussion of relevant implications
for the interpretation of quantum theory.

\section{Einstein, Podolsky, Rosen, and Bell\label{sec:Einstein-Podolsky-Rosen-and-Bell}}

The EPR argument~\citep{EinsteinPodolskyRosen:1935cqmdprbcc} was
based on a rudimentary version of \emph{quantum steering}, a term
introduced by Erwin Schr{\"o}dinger shortly thereafter~\citep{Schrodinger:1935dprbss,Schrodinger:1936prbss}. 

In quantum steering, two observers, Alice and Bob, split a pair of
quantum systems described by an entangled wave function, and then
move a large distance apart. If Bob decides to carry out a local measurement
on his system, then his choice of measurement basis will appear to
\textquoteleft steer\textquoteright{} Alice's system to collapse to
a corresponding basis. However, Bob will not be able to control which
specific wave function Alice's system selects in that basis, nor will
Alice be aware that anything strange has happened until she later
confers with Bob. (Note that this paper will use the terms \textquoteleft wave
function\textquoteright{} and \textquoteleft state vector\textquoteright{}
interchangeably.) 

Nonetheless, the overall behavior of the entangled pair of systems
looks suspiciously like a form of causal nonlocality\textemdash a
concrete manifestation of what Einstein in 1947 called ``spooky action
at a distance'' (``\emph{spukhafte Fernwirkung}'')~\citep{Einstein:1947ltmb}.
Following the EPR paper's publication, Schr{\"o}dinger described the situation
in the following way: 
\begin{quotation}
It is rather discomforting that the theory should allow a system to
be steered or piloted into one or the other type of state at the experimenter's
mercy in spite of his having no access to it.~\citep{Schrodinger:1935dprbss}
\end{quotation}

The EPR paper took for granted that causal nonlocality should be impossible,
and argued that the only available alternative was to assert that
the faraway system should already know what measurement result it
would reveal according to any hypothetical choice of measurement basis.
Because this information was not encoded in the system's overall wave
function, the authors of the EPR paper concluded that quantum theory
was incomplete. Indeed, the EPR paper was titled ``Can {[}the{]}
Quantum-Mechanical Description of Physical Reality Be Considered Complete?''

If one were to regard the EPR paper's reasoning as sound, then one
would seemingly be confronted with the following logical fork: either
accept causal nonlocality in quantum theory, or instead assert both
the incompleteness of quantum theory and the existence of a causally
local way for measurement results to be ``predetermined,'' in the
language of John Bell's 1964 paper ``On the Einstein-Podolsky-Rosen
Paradox''~\citep{Bell:1964oeprp}. Writing about the EPR argument
in a 1981 paper, Bell described this logical fork in the following
way: 
\begin{quotation}
For after observing only one particle{[},{]} the result of subsequently
observing the other (possibly at a very remote place) is immediately
predictable. Could it be that the first observation somehow fixes
what was unfixed, or makes real what was unreal, not only for the
near particle{[},{]} but also for the remote one? For EPR{[},{]} that
would be an unthinkable \textquoteleft spooky action at a distance.\textquoteright{}
To avoid such action at a distance{[},{]} they have to attribute,
to the space-time {[}sic{]} regions in question, \emph{real} properties
in advance of observation, correlated properties, which \emph{predetermine}
the outcomes of these particular observations. Since these real properties,
fixed in advance of observation, are not contained in {[}the{]} quantum
formalism, that formalism for EPR is \emph{incomplete}.'' {[}Emphasis
in the original.{]}~\citep{Bell:1981bssatnor}
\end{quotation}

In his 1964 paper~\citep{Bell:1964oeprp}, Bell argued that this
logical fork was, in the end, a mirage, and that quantum theory unavoidably
entailed causal nonlocality. To set up his argument, Bell considered
general reformulations of quantum theory involving \textquoteleft hidden
variables\textquoteright{} that uniquely predetermined measurement
outcomes. Bell's goal was to show that any such \emph{measurement-deterministic}
hidden-variables theory would have to involve causally nonlocal effects.

As Bell noted in his 1964 paper, one such measurement-deterministic
hidden-variables theory was already known, at least for the case of
nonrelativistic systems of finitely many particles. Called the \emph{de Broglie-form pilot-wave formulation}
of quantum theory, or \emph{Bohmian mechanics}~\citep{deBroglie:1930iswm,Bohm:1952siqtthvi,Bohm:1952siqtthvii},
this theory featured faster-than-light action at a distance, which
Bell called ``a grossly nonlocal structure.'' 

The result of Bell's 1964 paper was the first version of what is now
called \emph{Bell's theorem}, which implied that if a measurement-deterministic
hidden-variables theory were based on \emph{causally local} dynamics,
according to Bell's criteria, then the theory should satisfy an inequality
that is violated in quantum theory. The 2022 Nobel Prize in Physics~\citep{AspectClauserZeilinger:2022tnpip2}
was awarded to Alain Aspect, John Clauser and Anton Zeilinger for
their experiments verifying that quantum systems indeed violate Bell's
inequality, fully in accord with the predictions of quantum theory.

Importantly, Bell's 1964 paper assumed the soundness of the EPR argument,
which, in turn, implicitly relied on several contestable principles.
These included appealing to an explicit form of wave-function collapse,
as well as treating measurement interventions as primitive axiomatic
ingredients of quantum theory.

At an even deeper level, the EPR argument depended on an \emph{interventionist conception of causation},
in which causation is supposed to be explicated in terms of abstract
\emph{agents} carrying out formal \emph{interventions} on one set
of variables that then imply changes in another set of variables.
(For a review of interventionist accounts of causation, see~\citep{Woodward:2023cam}.)
It is not obvious how to express the EPR argument more fundamentally
in terms of the constituent atoms that make up measuring devices and
embodied observers, all undergoing some global physical process. Nor
is it clear that the EPR argument would be applicable to any formulation
of quantum theory that foregoes not only primitive measurement interventions,
but also lacks unique measurement outcomes, such as Hugh Everett's
\textquoteleft many worlds\textquoteright{} interpretation~\citep{Everett:1957rsfqm,Everett:1973tuwf,DeWitt:1970qmr}.

Given these substantive reasons for doubting the EPR argument, Bell's
1964 results could not be taken to imply that quantum theory \emph{necessarily}
involved causal nonlocality. His 1964 results instead reduced to the
more modest consequence of only ruling out measurement-deterministic
hidden-variables theories obeying causally local dynamics.

Putting aside several other potential loopholes (see~\citep{MyrvoldGenoveseShimony:2024bst}
for a review), Bell's 1964 paper therefore left open three possibilities:
measurement-deterministic hidden-variables theories with \emph{nonlocal}
dynamics, hidden-variables theories with \emph{stochastic} measurement
outcomes, and formulations of quantum theory that eschewed hidden
variables altogether.

In 1975~\citep{Bell:1975ttolb}, Bell updated his theorem to encompass
the second and third of these classes of possibilities, where the
third class includes \textquoteleft textbook\textquoteright{} quantum
theory itself. (For pedagogical reviews of textbook quantum theory,
see~\citep{Sakurai:1993mqm,Shankar:1994pqm,GriffithsSchroeter:2018iqm}.)
Crucially, extending his theorem in this way required introducing
a controversial new criterion for causal locality, a principle that
Bell called ``local causality.'' Bell was able to show that all
formulations of quantum theory satisfying his principle of local causality
should obey a generalization of his inequality originally derived
in 1969 by John Clauser, Michael Horne, Abner Shimony, and Richard
Holt~\citep{ClauserHorneShimonyHolt:1969pettlhvt}. This generalized
inequality is likewise violated by quantum theory.

In keeping with the terminology of~\citep{MyrvoldGenoveseShimony:2024bst},
this paper will distinguish \textquoteleft local causality\textquoteright{}
from the more basic condition of \textquoteleft causal locality\textquoteright{}
defined in \eqref{eq:DefCausalityLocality}. In short, \textquoteleft causal
locality\textquoteright{} means that any causal influences that \emph{happen
to occur} in a given scenario should not propagate faster than light,
whereas \textquoteleft local causality\textquoteright{} \emph{positively}
\emph{asserts the existence} of local causal relationships in specific
situations.

There are several incorrect ways to read the stronger 1975 version
of Bell's theorem. One is that the theorem rules out hidden variables
altogether. Another false reading is that one can avoid violating
Bell's principle of local causality merely by avoiding the introduction
of hidden variables\textemdash but this reading confuses the weaker
1964 version of Bell's theorem with the stronger 1975 version, which
applies even to theories that do not include hidden variables at all,
like textbook quantum theory itself. The correct reading of Bell's
theorem is to stay close to what Bell himself wrote and conclude that
his principle of local causality is violated by all empirically adequate
formulations of quantum theory, including the textbook version of
the theory, again putting aside various potential loopholes.

It is far from clear, however, that the principle of local causality
that Bell used to prove the stronger version of his theorem was the
correct way to formulate the more basic condition of causal locality
in the first place. Bell himself warned against taking his principle
of local causality too seriously. Indeed, in a 1990 lecture~\citep{Bell:1990lnc},
he cautioned that his principle ``should be viewed with the utmost
suspicion.''

Bell had good reasons for being skeptical of his own theorem's premises,
due to his history with an older theorem proved by John von Neumann
decades before. That earlier theorem had been widely viewed as \emph{completely}
ruling out the possibility of hidden variables~\citep{vonNeumann:1927wadq,vonNeumann:1932mgdq,vonNeumann:2018mfoqmne}.
 Already in 1935, Grete Hermann had determined that von Neumann's
theorem depended on an assumption about expectation values that was
too narrow~\citep{hermann1935circularity,Seevinck:2017ctgghovnsnhvp}.
Bell essentially discovered the same flaw in von Neumann's proof decades
later~\citep{Bell:1966otpohviqm}. (For an excellent historical discussion
of von Neumann's theorem, its shortcomings, and its critics, see~\citep{Bacciagaluppi:2021tsibhavn12}.)

\section{Bell's Principle of Local Causality\label{sec:Bell's-Principle-of-Local-Causality}}

To lay the groundwork for the discussion ahead, it will be important
to begin with a brief presentation of the 1964 and 1975 versions of
Bell's theorem, with a focus on their key implicit assumptions. It
is precisely these implicit assumptions that will be challenged in
this paper,  for the eventual purpose of developing a better criterion
for causal locality.

In his 1990 lecture, Bell noted the limitations of textbook quantum
theory, which lacked any notion of ``local \emph{be}ables''\textemdash meaning
\emph{actual} properties possessed by localized physical systems\textemdash as
opposed to the theory's more austere and instrumentalist notions of
\emph{observables}, \emph{measurement settings}, and \emph{measurement outcomes}:
\begin{quotation}
Even then, we are frustrated by the vagueness of contemporary quantum
mechanics. You will hunt in vain in the text-books {[}sic{]} for the
local \emph{be}ables of the theory. What you may find there are the
so-called \textquoteleft local observables\textquoteright . It is
then implicit that the apparatus of \textquoteleft observation\textquoteright ,
or, better, of experimentation, and the experimental results, are
real and localized. We will have to do as best we can with these rather
ill-defined local beables, while hoping always for a more serious
reformulation of quantum mechanics where the local beables are explicit
and mathematical rather than implicit and vague. {[}Emphasis in the
original.{]}~\citep{Bell:1990lnc}
\end{quotation}

In setting up the 1964 version of his theorem~\citep{Bell:1964oeprp},
Bell resorted to a pair of bivalent  measurement outcomes $A=\pm1$
and $B=\pm1$ at far separation in space, together with their respective
local measurement settings $\vec a$ and $\vec b$, with the special
feature that if $\vec a=\vec b$, then $A=-B$. Bell then imagined
a measurement-deterministic hidden-variables theory containing a set
of hidden variables $\lambda$, and supposed that these hidden variables
$\lambda$, together with the measurement settings $\vec a$ and $\vec b$,
fully predetermined the values of the measurement outcomes $A$ and
$B$: 
\begin{equation}
A=A\left(\vec a,\vec b,\lambda\right)=\pm1,\quad B=B\left(\vec a,\vec b,\lambda\right)=\pm1.\label{eq:1964OutcomeDeterminism}
\end{equation}
Following the terminology of~\citep{MyrvoldGenoveseShimony:2024bst},
this assumption will be called \emph{Outcome Determinism}.

In that 1964 paper, Bell's causal-locality assumptions included the
condition that the measurement outcome $A$ should not depend on the
faraway measurement setting $\vec b$, and, similarly, that the measurement
outcome $B$ should not depend on the faraway measurement setting
$\vec a$. Bell concluded that $A$ should be a function $A\left(\vec a,\lambda\right)$
of $\vec a$ and $\lambda$ alone, and that $B$ should be a function
$B\left(\vec b,\lambda\right)$ of $\vec b$ and $\lambda$ alone:
\begin{equation}
A\left(\vec a,\vec b,\lambda\right)=A\left(\vec a,\lambda\right),\quad B\left(\vec a,\vec b,\lambda\right)=B\left(\vec b,\lambda\right).\label{eq:1964ParameterIndependence}
\end{equation}
 Today these assumptions are known as \emph{Parameter Independence}~\citep{Shimony:1986eapitqw}.

Crucially, Bell's proof also relied on a special implication of Outcome
Determinism and Parameter Independence. Letting $\rho\left(\lambda\right)$
denote an assumed probability distribution for the hidden variables,
Outcome Determinism \eqref{eq:1964OutcomeDeterminism} and Parameter
Independence \eqref{eq:1964ParameterIndependence} suggested that
the expectation value of the product of the measurement outcomes $A$
and $B$ over many runs of the experiment should be given by 
\begin{equation}
P\left(\vec a,\vec b\right)=\int d\lambda\,\rho\left(\lambda\right)A\left(\vec a,\lambda\right)B\left(\vec b,\lambda\right).\label{eq:Bell1964ProductExpectationValue}
\end{equation}
 (As an aside, notice that the very existence of the probability distribution
$\rho\left(\lambda\right)$ for the hidden variables was yet one more
implicit assumption in Bell's proof.)

Invoking the formula \eqref{eq:Bell1964ProductExpectationValue} for
the expectation value $P\left(\vec a,\vec b\right)$, the end-result
of the 1964 paper was the well-known \emph{Bell inequality}: 
\begin{equation}
1+P\left(\vec b,\vec c\right)\geq\absval{P\left(\vec a,\vec b\right)-P\left(\vec a,\vec c\right)}.\label{eq:Bell1964Inequality}
\end{equation}
 Here $\vec c$ is an alternative choice of measurement setting.
Quantum theory predicts violations of this inequality, and, again,
the 2022 Nobel Prize in Physics~\citep{AspectClauserZeilinger:2022tnpip2}
was awarded for the experimental confirmation of those violations. 

Given Bell's criticism~\citep{Bell:1966otpohviqm} of von Neumann's
hidden-variables theorem over its assumptions about expectation values,
as described earlier in this paper, it is ironic that Bell's own theorem
likewise hinged on a statement about how expectation values were supposed
to work. Without Outcome Determinism and Parameter Independence, the
formula \eqref{eq:Bell1964ProductExpectationValue} is not the correct
way to calculate the necessary expectation value. 

To see why, consider a theory with stochastic measurement outcomes,
as in Bell's 1975 paper, with some set of variables $\lambda$ representing
beables, whether hidden variables or not. (As noted by Bell in \citep{Bell:1981bssatnor},
one could even try to regard wave functions themselves as \textquoteleft spatially
nonseparable beables.\textquoteright ) For this more general case,
in the formula \eqref{eq:Bell1964ProductExpectationValue} for the
expectation value, one then needs to replace the product 
\begin{equation}
A\left(\vec a,\lambda\right)B\left(\vec b,\lambda\right)\label{eq:1964BellProductMeasurementOutcomes}
\end{equation}
 with the statistical average 
\begin{equation}
\sum_{A,B}\rho\left(A,B\given\vec a,\vec b,\lambda\right)AB,\label{eq:1975BellProductMeasurementOutcomesAsAverage}
\end{equation}
 where $\rho\left(A,B\given\vec a,\vec b,\lambda\right)$ is some
joint probability distribution conditioned on the measurement settings
$\vec a$ and $\vec b$, as well as conditioned on the variables $\lambda$
representing the theory's beables. It follows that \eqref{eq:Bell1964ProductExpectationValue}
should be replaced with 
\begin{equation}
P\left(\vec a,\vec b\right)=\int d\lambda\,\rho\left(\lambda\right)\sum_{A,B}\rho\left(A,B\given\vec a,\vec b,\lambda\right)AB.\label{eq:1975BellProductExpectationValue}
\end{equation}
 In place of Outcome Determinism \eqref{eq:1964OutcomeDeterminism}
and Parameter Independence \eqref{eq:1964ParameterIndependence},
one then needs new assumptions in order to derive something like the
Bell inequality \eqref{eq:Bell1964Inequality}.

From the standard rules for working with conditional probabilities,
one can always write down the decomposition 
\begin{equation}
\rho\left(A,B\given\vec a,\vec b,\lambda\right)=\rho\left(A\given\vec a,\vec b,\lambda,B\right)\rho\left(B\given\vec a,\vec b,\lambda\right).\label{eq:1975BellFirstJointProbabilityFactorization}
\end{equation}
For a given measurement-stochastic theory, Bell's new \emph{principle of local causality}
was the condition that the theory should contain variables $\lambda$
representing a sufficiently rich collection of beables localized in
the overlap of the past light cones of the measurement outcomes $A$
and $B$ that $\lambda$ screens off $B$ and $\vec b$ from $A$,
and also screens off $\vec a$ from $B$, in the sense that 
\begin{equation}
\rho\left(A\given\vec a,\vec b,\lambda,B\right)=\rho\left(A\given\vec a,\lambda\right),\quad\rho\left(B\given\vec a,\vec b,\lambda\right)=\rho\left(B\given\vec b,\lambda\right).\label{eq:1975BellPrincipleOfLocalCausalityAsScreenOffCondition}
\end{equation}

Looking back at the decomposition \eqref{eq:1975BellFirstJointProbabilityFactorization},
it is clear that this new assumption \eqref{eq:1975BellPrincipleOfLocalCausalityAsScreenOffCondition}
is \emph{equivalent} to requiring that conditioning on the variables
$\lambda$ representing beables localized in the overlap of the past
light cones of $A$ and $B$ leads to the following factorization
condition: 
\begin{equation}
\rho\left(A,B\given\vec a,\vec b,\lambda\right)=\rho\left(A\given\vec a,\lambda\right)\rho\left(B\given\vec b,\lambda\right).\label{eq:1975BellPrincipleOfLocalCausalityAsFactorization}
\end{equation}
 Indeed, in his 1981 paper~\citep{Bell:1981bssatnor}, Bell took
this latter formula to be his basic principle of local causality,
and attempted to justify it on its own merits.

The factorization version \eqref{eq:1975BellPrincipleOfLocalCausalityAsFactorization}
of Bell's principle of local causality is, in turn, \emph{also} equivalent
to the conjunction of two other assumptions.

The first assumption is the following \emph{weaker} factorization
condition: 

\begin{equation}
\rho\left(A,B\given\vec a,\vec b,\lambda\right)=\rho\left(A\given\vec a,\vec b,\lambda\right)\rho\left(B\given\vec a,\vec b,\lambda\right).\label{eq:1975OutcomeIndependence}
\end{equation}
 This property is now called \emph{Outcome Independence}~\citep{Shimony:1986eapitqw}.

The other assumption is a generalization of Parameter Independence
\eqref{eq:1964ParameterIndependence} to mean that the conditional
probabilities for the measurement outcome $A$ do not depend on the
measurement setting $\vec b$, and that the conditional probabilities
for the measurement outcome $B$ do not depend on the measurement
setting $\vec a$: 
\begin{equation}
\rho\left(A\given\vec a,\vec b,\lambda\right)=\rho\left(A\given\vec a,\lambda\right),\quad\rho\left(B\given\vec a,\vec b,\lambda\right)=\rho\left(B\given\vec b,\lambda\right).\label{eq:1975ParameterIndependence}
\end{equation}

Assuming Outcome Independence \eqref{eq:1975OutcomeIndependence}
together with the updated version of Parameter Independence \eqref{eq:1975ParameterIndependence},
one obtains Bell's factorization \eqref{eq:1975BellPrincipleOfLocalCausalityAsFactorization},
where again $\lambda$ denotes variables representing a sufficiently
rich collection of beables localized in the overlap of the past light
cones of the measurement results $A$ and $B$. The expectation value
\eqref{eq:1975BellProductExpectationValue} then becomes
\begin{equation}
P\left(\vec a,\vec b\right)=\int d\lambda\,\rho\left(\lambda\right)\left(\sum_{A}\rho\left(A\given\vec a,\lambda\right)A\right)\left(\sum_{B}\rho\left(B\given\vec b,\lambda\right)B\right),\label{eq:1975BellProductExpectationValueWithOIAndPISimplified}
\end{equation}
 which closely resembles the 1964 version \eqref{eq:Bell1964ProductExpectationValue}
of the same expectation value. This formula thereby makes it possible
to derive a more general form of the Bell inequality, as first obtained
in 1969 by Clauser, Horne, Shimony, and Holt~\citep{ClauserHorneShimonyHolt:1969pettlhvt}.
This inequality is violated by all theories that are empirically equivalent
to textbook quantum theory\textemdash including the textbook theory
itself\textemdash so all such theories must also violate Bell's principle
of local causality.

Bell's principle of local causality\textemdash in either of its equivalent
forms \eqref{eq:1975BellPrincipleOfLocalCausalityAsScreenOffCondition}
or \eqref{eq:1975BellPrincipleOfLocalCausalityAsFactorization}\textemdash implicitly
depends on an assumption that goes beyond questions of locality.
That implicit assumption is called \emph{Reichenbach's principle of common causes}.
(For a review, see Section~19 of Hans Reichenbach's book~\citep{Reichenbach:1956tdot},
and also \citep{HitchcockRedei:2021rsccp}.)  

Reichenbach's principle of common causes states that if two variables
$A$ and $B$ are correlated, in the sense that their joint probability
$P\left(A,B\right)$ fails to factorize as the product of their standalone
probabilities $P\left(A\right)$ and $P\left(B\right)$, 
\begin{equation}
P\left(A,B\right)\ne P\left(A\right)P\left(B\right),\label{eq:CorrelationForReichenbach}
\end{equation}
 and if $A$ and $B$ do not causally influence each other, then there
should exist some other variable $C$ such that conditioning on $C$
leads to the following factorization: 
\begin{equation}
P\left(A,B\given C\right)=P\left(A\given C\right)P\left(B\given C\right).\label{eq:ReichenbachCommonCausePrincipleFactorization}
\end{equation}
 That is, Reichenbach's principle \emph{positively asserts the existence}
of a \textquoteleft common-cause\textquoteright{} variable $C$ for
$A$ and $B$. In this way, the variable $C$ is said to \textquoteleft explain\textquoteright{}
or \textquoteleft account for\textquoteright{} the correlation between
$A$ and $B$.\footnote{Note that this presentation of the principle is slightly generalized
from Reichenbach's original formulation, which assumed that $A$ and
$B$ were \emph{positively} correlated, so that $P\left(A,B\right)>P\left(A\right)P\left(B\right)$.}

Bell's principle of local causality\textemdash again in either of
its equivalent forms \eqref{eq:1975BellPrincipleOfLocalCausalityAsScreenOffCondition}
or \eqref{eq:1975BellPrincipleOfLocalCausalityAsFactorization}\textemdash clearly
invokes Reichenbach's principle, with the role of the asserted common-cause
variable $C$ played by the variables $\lambda$ representing beables
localized in the overlap of the past light cones of the measurement
results $A$ and $B$.

Reichenbach's principle of common causes may seem sensible and intuitive
in the context of everyday experience, but those are far from definitive
reasons to take it to be a fundamental requirement for causal locality.
In particular, embedded in both Reichenbach's principle of common
causes and Bell's principle of local causality is the assumption that
the asserted common causes in question must specifically take the
form of variables that can be conditioned on and then summed or integrated
over.

Just as a formulation of quantum theory that violates von Neumann's
assumptions about expectation values can evade von Neumann's theorem
and thereby admit hidden variables, a formulation of quantum theory
that fails to adhere to the strictures of Reichenbach's principle
of common causes could violate Bell's principle of local causality
without necessarily entailing nonlocal causation\textemdash as was
pointed out, for example, by William Unruh:
\begin{quotation}
It is true that this common cause cannot be stated in exactly the
form which for example Reichenbach set up to describe common causes
for a classical statistical system. But that is not surprising. Quantum
mechanics is not classical mechanics. The structure of the correlations
in a quantum system differ from those in a classical system, as Bell
so succinctly showed. But those correlations do not arise mysteriously
somehow in the development of a widely spaced system. Those correlations
do not require some mysterious non-local {[}sic{]} action to be explained.
They are simply there, as are correlations in a classical system,
due to the evolution from a common (quantum) cause in the past.~\citep{Unruh:2002iqmnl}
\end{quotation}

Returning once again to Bell's 1990 lecture~\citep{Bell:1990lnc},
Bell actually formulated \emph{two} versions of his principle of local
causality.

Bell identified the first version as the following statement:

\begin{equation}
\left.\begin{minipage}{\columnwidth}
\leftskip=10pt 
\rightskip=60pt 

The direct causes (and effects) of events are near by {[}sic{]}, and
even the indirect causes (and effects) are no further away than permitted
by the velocity of light.

\end{minipage}\hspace{-50pt}\right\}
\label{eq:BellFirstPrincipleLocalCausality}
\end{equation}

{\noindent}This first version is very close in spirit to the condition
of causal locality introduced at the beginning of this paper in \eqref{eq:DefCausalityLocality},
and is merely a locality condition on whatever causal influences \emph{happen
to occur}.

However, Bell then stated that ``The above principle of local causality
is not yet sufficiently sharp and clean for mathematics,'' followed
by ``Now it is precisely in cleaning up intuitive ideas for mathematics
that one is likely to throw out the baby with the bathwater. So the
next step should be viewed with the utmost suspicion.'' It was at
this point that Bell turned to the second version of his principle
of local causality, which \emph{positively asserted the existence}
of common causes and became the mathematical statement \eqref{eq:1975BellPrincipleOfLocalCausalityAsScreenOffCondition}.

This paper is hardly the first written argument to claim that Bell's
principle of local causality is not the correct way to capture causal
locality in a formulation of quantum theory. Beyond implicitly depending
on Reichenbach's principle of common causes, one should also note
that some readings of Bell's theorem, like several related theorems~\citep{ClauserHorneShimonyHolt:1969pettlhvt,GreenbergerHorneZeilinger:1989gbbt},
assume a notion of causation based on treating measurement settings
and measurement results as primitive interventions by abstract agents.
That is, these theorems depend on an interventionist conception of
causation, as defined earlier in this paper. It is therefore not clear
whether the theorems would make sense if one were instead to work
at the level of the constituent atoms of the relevant measuring devices
and physically embodied observers, all as parts of some sort of global
probabilistic process.

Indeed, when thinking in terms of a global probabilistic process,
without abstract agents and primitive interventions, it is far from
obvious how to identify causal influences or even nonlocal interactions,
especially without concrete notions like Newtonian forces that are
capable of establishing definitive physical linkages between systems.
(For an introduction to some of the challenges that arise when attempting
to make sense of causation in physics, see~\citep{Frisch:2024cip}.)

Other theorems, such as~\citep{BongUtrerasAlarconGhafariLiangTischlerCavalcantiPrydeWiseman:2020asngtotwfp},
depend on strong assumptions about the existence of theoretical joint
probability distributions involving the measurement results of subsystems
at \emph{intermediate} times during an overall unitary process. The
new formulation of quantum theory to be reviewed shortly provides
principled reasons why such theoretical joint probability distributions
should not be assumed to exist\textemdash the formulation simply does
not supply them in its microphysical laws, due in part to \emph{indivisibility},
a concept that will turn out to play a central role. 

\section{The Unistochastic Formulation of Quantum Theory\label{sec:The-Unistochastic-Formulation-of-Quantum-Theory}}

As described in \citep{Barandes:2023tsqc,Barandes:2023tsqt}, one
can reformulate quantum theory in terms of a sufficiently general
theory of stochastic processes, working entirely outside the traditional
\textquoteleft wave-function paradigm.\textquoteright{} Note that
this approach is not continuous with older attempts to formulate quantum
theory in stochastic terms~\citep{Bopp:1947qsuk,Bopp:1952efdqbsdk,Bopp:1953sudgdqde,Fenyes:1952ewbuidq,Nelson:1966dtobm,Nelson:1985qf},
all of which assumed a fundamental Markov condition, nor is it connected
with stochastic-collapse approaches to quantum theory~\citep{GhirardiRiminiWeber:1986udmms},
which treat wave functions or density matrices as basic ingredients
of physical reality.

The necessary axioms for this stochastic formulation are much simpler
and more transparent than for traditional textbook treatments of quantum
theory, without any need for metaphysically opaque postulates about
wave functions in abstract Hilbert spaces over the complex numbers.\footnote{Technically speaking, the Hilbert spaces of quantum theory are defined
not over the complex numbers alone, but over the \emph{pseudo-quaternions}~\citep{Stueckelberg:1960qtirhs},
which are a Clifford algebra generated by $1$, the imaginary unit
$i$, and the complex-conjugation operator $K$. This operator $K$
is needed for implementing time-reversal transformations, and satisfies
$K^{2}=1$ together with the anticommutation relation $Ki=-iK$. Altogether,
the elementary pseudo-quaternions $1$, $i$, $K$, and $iK$ satisfy
the basic relations $-i^{2}=K^{2}=\left(iK\right)^{2}=\left(i\right)\left(K\right)\left(iK\right)=1$.}

At the level of kinematics, one assumes a system with a set of configurations,
forming an old-fashioned configuration space $\configspace$. The
specific choice of configuration space depends on the particular kind
of system one is modeling, just like in classical physics, so $\configspace$
could consist of arrangements of particle positions, or of local field
intensities, or of digital bits, or of some other physical ingredients
altogether. 

Sticking for simplicity to the discrete case, perhaps after a suitable
degree of coarse-graining, the configuration space then consists of
a collection of configurations $i=1,\dots,N$. (One can generalize
the analysis ahead to the continuous case by introducing a measure
on the configuration space and by replacing summations with integrations.)

At the level of dynamics, the microphysical laws consist of conditional
or transition probabilities of the form 
\begin{equation}
\stochasticmatrix_{ij}\left(t\right)\defeq p\left(i,t\given j,0\right)\quad\left[\textrm{for }i,j=1,\dots N\right],\label{eq:TransitionMatrixAsConditionalProbabilities}
\end{equation}
 each of which supplies the probability for the system to be in its
$i$th configuration at a continuously variable time $t$, given that
the system is in its $j$th configuration at a suitable initial time
$0$. (No assumption is made here that $t>0$ or $t<0$.) Introducing
standalone probability distributions at the initial time $0$ and
at arbitrary times $t$, 
\begin{equation}
p_{j}\left(0\right)\defeq p\left(j,0\right),\quad p_{i}\left(t\right)\defeq p\left(i,t\right)\quad\left[\textrm{for }i,j=1,\dots N\right],\label{eq:StandaloneProbabilityDistributions}
\end{equation}
 the conditional or transition probabilities \eqref{eq:TransitionMatrixAsConditionalProbabilities}
that make up the basic microphysical laws give a simple linear relationship
between the standalone probabilities $p\left(j,0\right)$ at the initial
time $0$ and the standalone probabilities $p\left(i,t\right)$ at
the final time $t$, in accordance with the standard rules for conditional
probabilities and marginalization: 
\begin{equation}
p\left(i,t\right)=\sum_{j=1}^{N}p\left(i,t\given j,0\right)p\left(j,0\right)\quad\left[\textrm{for }i=1,\dots N\right].\label{eq:BayesianUpdateInitialToFinalStandaloneProbabilities}
\end{equation}
 Following the somewhat more succinct notation introduced above, this
linear relationship becomes 
\begin{equation}
p_{i}\left(t\right)=\sum_{j=1}^{N}\stochasticmatrix_{ij}\left(t\right)p_{j}\left(0\right)\quad\left[\textrm{for }i=1,\dots N\right].\label{eq:BayesianUpdateInitialToFinalStandaloneProbabilitiesSimplerNotation}
\end{equation}

Working in terms of matrices, one can write the standalone probability
distributions here as $N\times1$ column vectors, 
\begin{equation}
p\left(0\right)\defeq\begin{pmatrix}p_{1}\left(0\right)\\
\vdots\\
p_{N}\left(0\right)
\end{pmatrix},\quad p\left(t\right)\defeq\begin{pmatrix}p_{1}\left(t\right)\\
\vdots\\
p_{N}\left(t\right)
\end{pmatrix},\label{eq:DefStandaloneProbabilitiesAsColumnVectors}
\end{equation}
 and one can write the collection of transition probabilities as an
$N\times N$ \emph{transition matrix}, 
\begin{equation}
\stochasticmatrix\left(t\right)\defeq\begin{pmatrix}\stochasticmatrix_{11}\left(t\right) & \stochasticmatrix_{12}\left(t\right)\\
\stochasticmatrix_{21}\left(t\right) & \ddots\\
 &  & \stochasticmatrix_{NN}\left(t\right)
\end{pmatrix}.\label{eq:DefTransitionMatrix}
\end{equation}
 One can then naturally express the basic linear relationship \eqref{eq:BayesianUpdateInitialToFinalStandaloneProbabilitiesSimplerNotation}
as an elementary matrix product: 
\begin{equation}
p\left(t\right)=\stochasticmatrix\left(t\right)p\left(0\right).\label{eq:BayesianUpdateInitialToFinalStandaloneProbabilitiesAsMatrixEq}
\end{equation}

The $N\times N$ transition matrix $\stochasticmatrix\left(t\right)$
consists of non-negative entries, and its columns each sum to $1$:
\begin{equation}
\left.\begin{aligned}\stochasticmatrix_{ij}\left(t\right)\geq0\quad\left[\textrm{for }i,j=1,\dots,N\right],\\
\sum_{i=1}^{N}\stochasticmatrix_{ij}\left(t\right)=1\quad\left[\textrm{for }j=1,\dots N\right].
\end{aligned}
\quad\right\} \label{eq:TransitionMatrixIsColumnStochastic}
\end{equation}
 Mathematically speaking, these properties identify $\stochasticmatrix\left(t\right)$
as a \emph{(column) stochastic matrix}.

An important concept here is the historically recent notion of \emph{divisibility}~\citep{WolfCirac:2008dqc,MilzModi:2021qspaqnp},
which is loosely related to the well-known \emph{Markov property}.
For a \emph{divisible} transition matrix $\stochasticmatrix\left(t\right)$
with a variable time $t$, and given an intermediate time $t^{\prime}$
between $0$ and $t$, there always exists a valid stochastic matrix
$\stochasticmatrix\left(t\from t^{\prime}\right)$ such that one can
\textquoteleft divide\textquoteright{} the dynamics from $0$ to $t$
into subintervals from $0$ to $t^{\prime}$, and then from $t^{\prime}$
to $t$, as ordinary matrix multiplication: 
\begin{equation}
\underbrace{\stochasticmatrix\left(t\right)}_{0\textrm{ to }t}=\underbrace{\stochasticmatrix\left(t\from t^{\prime}\right)}_{t^{\prime}\textrm{ to }t}\underbrace{\stochasticmatrix\left(t^{\prime}\right)}_{0\textrm{ to }t^{\prime}}.\label{eq:DivisibleTransitionMatrix}
\end{equation}

By contrast, for the kind of stochastic process that is equivalent
to a quantum system, the transition matrix will generically be \emph{indivisible},
meaning that no valid such stochastic matrix $\stochasticmatrix\left(t\from t^{\prime}\right)$
satisfying the divisibility property \eqref{eq:DivisibleTransitionMatrix}
will exist. A stochastic process based on a  potentially indivisible
transition matrix will be called a \emph{generalized stochastic system}
or \emph{process}.

An $N\times N$ matrix $\stochasticmatrix$ is called a \emph{unistochastic matrix}
if there exists a (generally non-unique) $N\times N$ unitary matrix
$\timeevop$ such that the individual entries of $\stochasticmatrix$
are each the modulus-squares of the corresponding entries of $\timeevop$:
\begin{equation}
\stochasticmatrix_{ij}=\absval{\timeevop_{ij}}^{2}\quad\left[\textrm{for }i,j=1,\dots,N\right].\label{eq:DefUnistochasticMatrix}
\end{equation}
 In \citep{Horn:1954dsmatdoarm}, Alfred Horn originally called such
matrices ``ortho-stochastic,'' but that term is now reserved for
the special case in which $\timeevop$ can be taken to be a real-orthogonal
matrix. The term ``unistochastic'' appears to have first been introduced
by Robert Thompson in \citep{Thompson:1989uln}.

Crucially, notice that the equalities appearing in \eqref{eq:DefUnistochasticMatrix}
hold \emph{entry-by-entry}. That is, $\stochasticmatrix$ is not given
by a simple matrix product like $\timeevop^{\adj}\timeevop$, which
would just give the identity matrix $\idmatrix$, due to the unitarity
of $\timeevop$. In particular, the overall relationship between $\stochasticmatrix$
and $\timeevop$ does not commute with matrix multiplication.

A generalized stochastic system with a unistochastic transition matrix
$\stochasticmatrix\left(t\right)$ will be called a \emph{unistochastic system}
or \emph{process}. As proved in \citep{Barandes:2023tsqt}, one can
always assume that a generalized stochastic system is, in fact, a
unistochastic system, by slightly enlarging or \emph{dilating} the
configuration space if necessary, and invoking the \emph{Stinespring dilation theorem}~\citep{Stinespring:1955pfoc}.
It therefore suffices to focus one's attention on unistochastic systems.

Reconstructing quantum theory from the set of unistochastic systems
is then an extended mathematical exercise.

Given the $N\times N$ unistochastic transition matrix $\stochasticmatrix\left(t\right)$
for a given unistochastic system, one starts by taking the quantum
system's unitary \emph{time-evolution operator} to be a (generally
not-uniquely) associated $N\times N$ time-dependent unitary matrix
$\timeevop\left(t\right)$: 
\begin{equation}
\stochasticmatrix_{ij}\left(t\right)=\absval{\timeevop_{ij}\left(t\right)}^{2}\quad\left[\textrm{for }i,j=1,\dots,N\right].\label{eq:UnistochasticTransitionMatrixFromUnitaryEntries}
\end{equation}

Unlike the underlying unistochastic transition matrix $\stochasticmatrix\left(t\right)$,
this unitary time-evolution operator $\timeevop\left(t\right)$ satisfies
a divisibility condition in the form of the usual composition law
 
\begin{equation}
\underbrace{\timeevop\left(t\right)}_{0\textrm{ to }t}=\underbrace{\timeevop\left(t\from t^{\prime}\right)}_{t^{\prime}\textrm{ to }t}\underbrace{\timeevop\left(t^{\prime}\right)}_{0\textrm{ to }t^{\prime}},\label{eq:UnitaryTimeEvOpCompositionLaw}
\end{equation}
 where the \emph{relative} time-evolution operator $\timeevop\left(t\from t^{\prime}\right)$
is defined by 
\begin{equation}
\timeevop\left(t\from t^{\prime}\right)\defeq\timeevop\left(t\right)\timeevop^{\adj}\left(t^{\prime}\right)\label{eq:DefRelativeTimeEvOp}
\end{equation}
 and is guaranteed to be unitary. The fact that modulus-squaring the
entries of a matrix, as in \eqref{eq:UnistochasticTransitionMatrixFromUnitaryEntries},
does not commute with matrix multiplication accounts for the failure
of $\stochasticmatrix\left(t\right)$ likewise to be divisible.

Indeed, if one attempts to define a unistochastic transition matrix
$\stochasticmatrix\left(t\from t^{\prime}\right)$ from $t^{\prime}$
to $t$ based on the relative time-evolution operator \eqref{eq:DefRelativeTimeEvOp},
\begin{equation}
\stochasticmatrix_{ij}\left(t\from t^{\prime}\right)\defeq\absval{\timeevop_{ij}\left(t\from t^{\prime}\right)}^{2},\label{eq:DefAttemptedRelativeTransitionMatrixFromUnitary}
\end{equation}
 then one ends up with a discrepancy between the \emph{actual-indivisible}
dynamical evolution $\stochasticmatrix\left(t\right)$ from $0$ to
$t$ and the \emph{nearest-divisible} dynamical evolution $\stochasticmatrix\left(t\from t^{\prime}\right)\stochasticmatrix\left(t^{\prime}\right)$:
\begin{equation}
\stochasticmatrix\left(t\right)\ne\stochasticmatrix\left(t\from t^{\prime}\right)\stochasticmatrix\left(t^{\prime}\right).\label{eq:DiscrepancyStochasticMatrices}
\end{equation}
 From the standpoint of regarding the quantum system as a unistochastic
system, the well-known \emph{interference effects} of quantum theory
merely reflect this discrepancy: 
\begin{equation}
\stochasticmatrix\left(t\right)-\stochasticmatrix\left(t\from t^{\prime}\right)\stochasticmatrix\left(t^{\prime}\right)\ne0\quad\left[\textrm{interference effects}\right].\label{eq:DefInterferenceFromDiscrepancyStochasticMatrices}
\end{equation}

Writing the initial standalone probability distribution $p_{j}\left(0\right)$
as the diagonal entries of an $N\times N$ initial \emph{density matrix}
$\densitymatrix\left(0\right)$ whose other entries are all $0$s,
\begin{equation}
\densitymatrix\left(0\right)\defeq\diag{p_{1}\left(0\right),\dots,p_{N}\left(0\right)}\defeq\begin{pmatrix}p_{1}\left(0\right) & 0\\
0 & \ddots\\
 &  & p_{N}\left(0\right)
\end{pmatrix},\label{eq:DefInitialDensityMatrixFromInitialStandaloneProbabilities}
\end{equation}
 the quantum system's density matrix at all other times is defined
by the usual similarity transformation given by the time-evolution
operator $\timeevop\left(t\right)$: 
\begin{equation}
\densitymatrix\left(t\right)\defeq\timeevop\left(t\right)\densitymatrix\left(0\right)\timeevop^{\adj}\left(t\right).\label{eq:DefTimeDependentDensityMatrixFromInitialDensityMatrix}
\end{equation}
 Observe that the resulting time-dependent density matrix $\densitymatrix\left(t\right)$
is not generally diagonal for times $t\ne0$.

Notice also that the famous linearity of the time evolution of quantum
theory, as exhibited by relationship between $\densitymatrix\left(t\right)$
and $\densitymatrix\left(0\right)$, is not a mystery, but ultimately
descends from the linearity of the basic relationship \eqref{eq:BayesianUpdateInitialToFinalStandaloneProbabilitiesSimplerNotation},
which again follows directly from the standard rules for conditional
probabilities and marginalization.

Assuming sufficient smoothness in $t$, so that \emph{Stone's theorem}
applies~\citep{Stone:1930ltihs}, one can define the system's self-adjoint
\emph{Hamiltonian} $H\left(t\right)$ as the infinitesimal generator
of time translations, 
\begin{equation}
H\left(t\right)\defeq i\hbar\frac{\partial\timeevop\left(t\right)}{\partial t}\timeevop^{\adj}\left(t\right)=H^{\adj}\left(t\right),\label{eq:DefHamiltonian}
\end{equation}
 in which case the system's density matrix $\densitymatrix\left(t\right)$
satisfies the \emph{von Neumann equation}, 
\begin{equation}
i\hbar\frac{\partial\densitymatrix\left(t\right)}{\partial t}=\bracks{H\left(t\right),\densitymatrix\left(t\right)},\label{eq:VonNeumannEq}
\end{equation}
 where the brackets denote the usual matrix commutator (not a Poisson
bracket): 
\begin{equation}
\bracks{X,Y}\defeq XY-YX.\label{eq:DefCommutator}
\end{equation}
If the system's density matrix $\densitymatrix\left(t\right)$ is
rank-one, then it can be factorized in terms of a complex-valued $N\times1$
\emph{state vector} or \emph{wave function} $\Psi\left(t\right)$,
\begin{equation}
\densitymatrix\left(t\right)=\Psi\left(t\right)\Psi^{\adj}\left(t\right)\quad\left[\textrm{if rank-one}\right],\quad\Psi\left(t\right)\defeq\begin{pmatrix}\Psi_{1}\left(t\right)\\
\vdots\\
\Psi_{N}\left(t\right)
\end{pmatrix},\label{eq:RankOneFactorizationForStateVector}
\end{equation}
 in which case the state vector $\Psi\left(t\right)$ evolves according
to the \emph{Schr{\"o}dinger equation}, 
\begin{equation}
i\hbar\frac{\partial\Psi\left(t\right)}{\partial t}=H\left(t\right)\Psi\left(t\right).\label{eq:SchrodingerEq}
\end{equation}
 It is notable that these familiar quantum-theoretic equations emerge
from an underlying stochastic process, which ultimately consists of
a system moving along some trajectory in a prosaic configuration space
according to (indivisible) stochastic transition probabilities. 

Observe that the state vector or wave function $\Psi\left(t\right)$
appears here as just a convenient piece of secondary, derived mathematics,
rather than as anything like a primary or fundamental physical object.
In the context of this overall stochastic picture, the wave function
is not a piece of ontological furniture, but instead encodes epistemic
information\textemdash the system's probabilities\textemdash as well
as nomological information\textemdash the system's unistochastic microphysical
dynamics.

Given a \emph{random variable} $A\left(t\right)$ on the system's
configuration space, meaning a spectrum of magnitudes $a_{1}\left(t\right),\dots,a_{N}\left(t\right)$
that depend on the system's configuration $i=1,\dots,N$ and that
generically also depend explicitly on the time $t$, the statistical
expectation value of $A\left(t\right)$ is defined as 
\begin{equation}
\expectval{A\left(t\right)}\defeq\sum_{i=1}^{N}a_{i}\left(t\right)p_{i}\left(t\right).\label{eq:DefExpectationValueRandomVariable}
\end{equation}
 In terms of the system's density matrix $\densitymatrix\left(t\right)$,
as defined in \eqref{eq:DefTimeDependentDensityMatrixFromInitialDensityMatrix},
and introducing a diagonal matrix $A\left(t\right)$ according to
\begin{equation}
A\left(t\right)\defeq\diag{a_{1}\left(t\right),\dots,a_{N}\left(t\right)}\defeq\begin{pmatrix}a_{1}\left(t\right) & 0\\
0 & \ddots\\
 &  & a_{N}\left(t\right)
\end{pmatrix},\label{eq:DefDiagonalMatrixForRandomVariable}
\end{equation}
 one can rewrite the expectation value \eqref{eq:DefExpectationValueRandomVariable}
in the equivalent form 
\begin{equation}
\expectval{A\left(t\right)}=\tr\left(A\left(t\right)\rho\left(t\right)\right),\label{eq:ExpectationValueRandomVariableFromTraceDensityMatrix}
\end{equation}
 which looks just like the standard formula from quantum theory.

Consider the special case in which $A=\projector_{i}$ is a rank-one
projector consisting of a matrix with a $1$ in its $i$th diagonal
entry and $0$s in all its other entries: 
\begin{equation}
\projector_{i}\defeq\mathrm{diag}\parens{0,\dots,0,\underset{\mathclap{\substack{\uparrow\\
i\textrm{th entry}
}
}}{1},0,\dots,0}.\label{eq:DefConfigurationProjectors}
\end{equation}
 It follows that if $\densitymatrix\left(t\right)$ is similarly rank-one,
in the sense of being factorizable according to \eqref{eq:RankOneFactorizationForStateVector}
in terms of a state vector $\Psi\left(t\right)$, then the expectation
value \eqref{eq:ExpectationValueRandomVariableFromTraceDensityMatrix}
reduces to the simplest version of the Born rule: 
\begin{equation}
p_{i}\left(t\right)=\absval{\Psi_{i}\left(t\right)}^{2}.\label{eq:SimplestBornRule}
\end{equation}

Random variables on the unistochastic system's configuration space
have the status of \emph{beables}, in Bell's terminology. By modeling
the measurement process in detail\textemdash treating measurement
devices as mundane stochastic systems in their own right\textemdash one
can show that non-diagonal self-adjoint operators represent observables
that are \emph{emergent phenomena} at the level of measurements,
and so are called \emph{emergeables} in \citep{Barandes:2023tsqc}.
A unistochastic system's beables and emergeables together comprise
the system's full noncommutative algebra of \emph{observables}.

Just as one can represent a stochastic process in the Hilbert-space
formalism familiar from quantum theory, one can take any quantum system
in its Hilbert-space formalism and turn the relationship \eqref{eq:UnistochasticTransitionMatrixFromUnitaryEntries}
around to define a corresponding stochastic process. This \emph{stochastic-quantum correspondence}
is a many-to-one relationship in both directions\textemdash a single
stochastic process will generally have many different-looking Hilbert-space
representations, and a given quantum system in its Hilbert-space formalism
may represent many different-looking stochastic processes. The relationship
between a stochastic process and its corresponding Hilbert-space representation
is therefore analogous to the relationship between a classical-deterministic
system described by second-order differential equations of motion
and its corresponding Hamiltonian phase-space representation, a relationship
that is likewise many-to-one in both directions.

At a practical level, one can therefore regard the Hilbert-space formalism
as a form of \textquoteleft analytical mechanics\textquoteright{}
for highly general stochastic processes, just as the Hamiltonian phase-space
formalism provides an analytical mechanics for a second-order classical-deterministic
system. Like any form of analytical mechanics, the Hilbert-space formalism
provides a powerful set of mathematical tools for specifying microphysical
laws in a systematic manner, for studying dynamical symmetries, for
proving theorems, and for calculating predictions.

The fact that one can reformulate a given quantum system as a unistochastic
system deflates much of the exotic talk about quantum theory. As spelled
out in~\citep{Barandes:2023tsqc}, from the standpoint of this \emph{unistochastic formulation}
of quantum theory, the \emph{measurement problem} arguably disappears,
because measuring devices are now to be modeled as ordinary (if complicated)
subsystems of an overall stochastic process, and one can show that
they end up in measurement-outcome configurations probabilistically
in accord with the usual predictions of the Born rule. Moreover, superposition
is no longer a literal smearing of configurations, interference is
just a breakdown \eqref{eq:DefInterferenceFromDiscrepancyStochasticMatrices}
in divisible dynamics, and decoherence is merely the leakage of statistical
correlations out into the larger environment. 

In particular, as explained in~\citep{Barandes:2023tsqc}, decoherence
automatically generates \emph{division events}, which are new times
$t^{\prime}$ at which the microphysical transition matrix $\stochasticmatrix\left(t\right)$
\emph{does} divide, in the sense of \eqref{eq:DivisibleTransitionMatrix}.
 A division event $t^{\prime}$ is therefore a time that can serve
in place of the initial time $0$ in the unistochastic system's microphysical
conditional probabilities.

If $t^{\prime}$ is a division event, then the unistochastic system
contains genuine microphysical conditional probabilities of the form
\begin{equation}
\stochasticmatrix_{ii^{\prime}}\left(t\from t^{\prime}\right)\defeq p\left(i,t\given i^{\prime},t^{\prime}\right),\label{eq:ConditionalProbabilityForDivisionEvent}
\end{equation}
 which are conditioned on the system's configuration $i^{\prime}$
at the division event $t^{\prime}$, where $\stochasticmatrix\left(t\from t^{\prime}\right)$
is a valid stochastic matrix satisfying the divisibility condition
\eqref{eq:DivisibleTransitionMatrix}. One expects that for a macroscopic
system in strong contact with a noisy environment that eavesdrops
on the system's configuration over a characteristic time scale $\delta t$,
the system's microphysical laws will become effectively Markovian
for time steps of duration $\delta t$.

\section{Bayesian Networks and Causation\label{sec:Bayesian-Networks-and-Causation}}

As explained earlier, the traditional textbook formulation of quantum
theory does not provide a hospitable domain for a non-interventionist
account of causation, making it very difficult to devise clear statements
about causal influences in general or causal locality in particular.
By replacing the Hilbert-space axioms with a true set of microphysical
laws consisting of conditional probabilities $\stochasticmatrix_{ij}\left(t\right)\defeq p\left(i,t\given j,0\right)$,
as introduced in \eqref{eq:TransitionMatrixAsConditionalProbabilities},
the new unistochastic formulation of quantum theory reviewed in this
paper opens up an important connection with the literature on \emph{Bayesian networks}~\citep{Pearl:2009cmrai},
which provide a much more amenable foundation for a non-interventionist
causal account.

In simple terms, a Bayesian network is a model that consists of a
set of random variables connected by a collection of conditional probabilities.
Displayed graphically, a Bayesian network will typically denote the
random variables by \emph{nodes}, and will denote the conditional
probabilities by directed line segments or \emph{edges} connecting
some of those nodes together.

\begin{figure}
\includegraphics[scale=0.6]{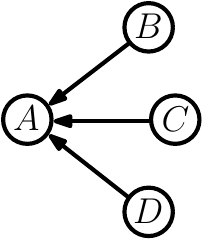}

\caption{\label{fig:SimpleFourVarBayesianNetwork}A simple Bayesian network
with four random variables $A$, $B$, $C$, and $D$ denoted by nodes,
with directed edges pointing to $A$ from $B$, $C$, and $D$.}

\end{figure}

For example, if a node representing a random variable $A$ is at the
pointed end of directed edges from nodes representing random variables
$B$, $C$, and $D$, as in Figure~\ref{fig:SimpleFourVarBayesianNetwork},
then the Bayesian network must supply a basic conditional probability
distribution $p\left(a\given b,c,d\right)$ among its laws, where
lowercase letters denote the possible values of the corresponding
random variables: 
\begin{equation}
p\left(a\given b,c,d\right)\defeq p\left(A=a\given B=b,C=c,D=d\right).\label{eq:DefBayesianNetworkFourVarDirectedConditionalProbability}
\end{equation}
 This conditional probability is the probability that the random variable
$A$ has the value $a$, given that the random variables $B$, $C$,
and $D$ have the respective values $b$, $c$, and $d$. In other
words, the values of $B$, $C$, and $D$ determine the conditional
probability distribution for the values of $A$.

It follows that if the random variables $B$, $C$, and $D$ were
to develop \emph{contingent} joint probabilities $p\left(b,c,d\right)$
in some concrete, real-life instantiation of the Bayesian network,
then the random variable $A$ would automatically inherit a contingent
standalone probability distribution $p\left(a\right)$ of its own
according to the standard multilinear rule 
\begin{equation}
p\left(a\right)=\sum_{b,c,d}p\left(a\given b,c,d\right)p\left(b,c,d\right).\label{eq:BayesianNetworkFourVarMultilinearRuleForProbability}
\end{equation}
 Said in another way, the basic conditional probabilities $p\left(a\given b,c,d\right)$,
together with the contingent joint probabilities $p\left(b,c,d\right)$
for $B$, $C$, and $D$, dictate the contingent standalone probabilities
$p\left(a\right)$ for $A$, and they do so in a multilinear way.

Importantly, the basic conditional probability distribution $p\left(a\given b,c,d\right)$
supplied by the Bayesian network in the present example is \emph{directed},
in the sense that the value $a$ of the random variable $A$ appears
to the \emph{left} of the \textquoteleft given\textquoteright{} symbol
$\given$, whereas the respective values $b$, $c$, and $d$ of the
random variables $B$, $C$, and $D$ appear to the \emph{right}.
To understand the significance of this directedness, it will be worthwhile
to construct a different conditional probability for comparison.

To that end, notice that if one were to combine the Bayesian network's
\emph{basic} conditional probability distribution $p\left(a\given b,c,d\right)$
with the \emph{contingent} joint probability distribution $p\left(b,c,d\right)$,
then one could formally define a joint probability distribution $p\left(a,b,c,d\right)$
for all four of the random variables $A$, $B$, $C$, and $D$ by
invoking the standard rule for conditional probabilities: 
\begin{equation}
p\left(a,b,c,d\right)\defeq p\left(a\given b,c,d\right)p\left(b,c,d\right).\label{eq:BayesianNetworkFourVarJointProbabilityFromConditional}
\end{equation}
 Defining a joint probability distribution $p\left(a,c,d\right)$
for $A$, $C$, and $D$ alone by marginalizing the joint probability
distribution $p\left(a,b,c,d\right)$ over $B$, 
\begin{equation}
p\left(a,c,d\right)\defeq\sum_{b}p\left(a,b,c,d\right),\label{eq:BayesianNetworkFourVarMarginalizeToThree}
\end{equation}
 and assuming that this joint probability $p\left(a,c,d\right)\ne0$
were nonzero, one could then formally condition on $A$, $C$, and
$D$ to obtain the conditional probability 
\begin{equation}
p\left(b\given a,c,d\right)\defeq\frac{p\left(a,b,c,d\right)}{p\left(a,c,d\right)}\quad\left[\textrm{if }p\left(a,c,d\right)\ne0\right].\label{eq:BayesianNetworkFourVarAlternativeConditional}
\end{equation}
 Writing out this conditional probability in more detail, one would
obtain the formula 
\begin{equation}
p\left(b\given a,c,d\right)=\frac{p\left(a\given b,c,d\right)p\left(b,c,d\right)}{\sum_{b^{\prime}}p\left(a\given b^{\prime},c,d\right)p\left(b^{\prime},c,d\right)},\label{eq:BayesianNetworkFourVarAlternativeConditionalDetail}
\end{equation}
 which makes clear that $p\left(b\given a,c,d\right)$ would depend
on the contingent joint probability distribution $p\left(b,c,d\right)$\textemdash and
in a nonlinear manner. Hence, although $p\left(b\given a,c,d\right)$
might exist, it would be a \emph{derived} conditional probability
distribution that depended on the contingencies of the given concrete
instantiation of the Bayesian network, and would therefore have a
different physical status from the \emph{basic, nomological} conditional
probability distribution $p\left(a\given b,c,d\right)$ supplied by
the Bayesian network's laws.

There exists a reading of a Bayesian network as a model of causal
relationships, with causal influences manifesting as the Bayesian
network's directed conditional probabilities. That is, if the Bayesian
network supplies a directed conditional probability distribution $p\left(a\given b,c,d\right)$
in its basic laws, then one should read the Bayesian network as implying
that the random variables $B$, $C$, and $D$ causally influence
the random variable $A$.

Although the causal influences encoded in Bayesian networks can be
given an interventionist cast, a non-interventionist interpretation
is available as well, with stochastic fluctuations in $B$, $C$,
and $D$ dictating stochastic fluctuations in $A$ through the directed
conditional probability distribution $p\left(a\given b,c,d\right)$.\footnote{Note that this conception of causation as corresponding to directed
conditional probability distributions is fundamentally distinct from
\emph{probability-raising theories of causation}. In particular,
no assumption is made here that the directed conditional probabilities
specifically raise any standalone probabilities.}

Notice how the directedness of the conditional probability distributions
supplied by a Bayesian network captures the inherently asymmetric
nature of cause-and-effect relationships.

Interestingly, this connection between the directedness of a Bayesian
network's basic conditional probabilities and the asymmetry of cause-and-effect
also sheds light on why causal language is so fraught in the context
of theories that are based on microphysical laws that are deterministic
and reversible. In a deterministically reversible theory, if a value
$a$ of a variable $A$ implies a corresponding value $b$ of another
variable $B$, then $p\left(b\given a\right)=1$, and, in addition,
any contingent standalone probability $p\left(a\right)$ assigned
to $a$ will necessarily equal the contingent standalone probability
$p\left(b\right)$ assigned to $b$. It follows immediately from Bayes'
theorem that $p\left(a\given b\right)=p\left(b\given a\right)=1$,
so these conditional probabilities are not directed, and the asymmetry
of cause-and-effect relationships is lost.

\section{A Microphysical Account of Causation\label{sec:A-Microphysical-Account-of-Causation}}

As reviewed in this paper, one can reformulate a quantum system in
terms of an underlying unistochastic system. The microphysical laws
of that unistochastic system consist of directed conditional probabilities
\eqref{eq:TransitionMatrixAsConditionalProbabilities}, $\stochasticmatrix_{ij}\left(t\right)\defeq p\left(i,t\given j,0\right)$,
which are very much like the directed conditional probabilities that
define the basic laws of a Bayesian network. Taking this resemblance
seriously, one can read the microphysical laws of the unistochastic
system as providing a microphysical notion of causal influences.

To make things more concrete, suppose that the unistochastic system
consists of two subsystems $Q$ and $R$, in the sense that $i=\left(q_{t},r_{t}\right)$
and $j=\left(q_{0},r_{0}\right)$, where lowercase letters denote
specific configurations of the corresponding subsystems. One can then
write the directed conditional probabilities \eqref{eq:TransitionMatrixAsConditionalProbabilities}
for the overall system as 
\begin{equation}
p\left(\left(q_{t},r_{t}\right),t\given\left(q_{0},r_{0}\right),0\right).\label{eq:PairSubsystemsDirectedConditionalProbability}
\end{equation}
 To say that the subsystem $Q$ is free of causal influences from
the subsystem $R$ over the time interval from $0$ to $t$ would
then be the statement that after marginalizing over the configuration
$r_{t}$ of $R$, the resulting conditional probability distribution
no longer depends on $r_{0}$: 
\begin{equation}
p\left(q_{t},t\given\left(q_{0},r_{0}\right),0\right)=p\left(q_{t},t\given q_{0},0\right).\label{eq:PairSubsystemsDirectedConditionalProbabilityNoCausalInfluence}
\end{equation}

\section{An Improved Principle of Causal Locality\label{sec:An-Improved-Principle-of-Causal-Locality}}

One can now formulate an improved \emph{principle of causal locality}:
\begin{equation}
\left.\begin{minipage}{\columnwidth}
\leftskip=10pt 
\rightskip=60pt 

A theory with microphysical directed conditional probabilities is
causally local if any pair of localized systems $Q$ and $R$ that
remain at spacelike separation for the duration of a given physical
process do not exert causal influences on each other during that process,
in the sense that the directed conditional probabilities for $Q$
are independent of $R$, and vice versa.

\end{minipage}\hspace{-50pt}\right\}
\label{eq:PrincipleCausalLocality}
\end{equation}

{\noindent}Having stated this new principle of causal locality,
one can show that quantum theory, formulated as a theory of unistochastic
processes, indeed satisfies it. 

For that purpose, consider a unistochastic system consisting of a
pair of localized subsystems $Q$ and $R$ that remain at spacelike
separation during a given physical process. The overall system's unistochastic
transition matrix $\stochasticmatrix_{QR}\left(t\right)$ has a corresponding
unitary time-evolution operator $\timeevop_{QR}\left(t\right)$ in
the sense of \eqref{eq:UnistochasticTransitionMatrixFromUnitaryEntries}.
Invoking the spacelike separation of $Q$ and $R$ together with the
usual assumptions employed in textbook quantum theory, the overall
time-evolution operator $\timeevop_{QR}\left(t\right)$ tensor-factorizes
into respective unitary time-evolution operators $\timeevop_{Q}\left(t\right)$
for $Q$ and $\timeevop_{R}\left(t\right)$ for $R$ individually:
\begin{equation}
\timeevop_{QR}\left(t\right)=\timeevop_{Q}\left(t\right)\tensorprod\timeevop_{R}\left(t\right).\label{eq:TimeEvOpParentSystemSubsystemsFactorizeSpacelikeSeparation}
\end{equation}

In contrast with matrix multiplication, tensor products \emph{do}
commute with modulus-squaring the entries of a matrix, so the overall
system's unistochastic transition matrix $\stochasticmatrix_{QR}\left(t\right)$
likewise tensor-factorizes: 
\begin{equation}
\stochasticmatrix_{QR}\left(t\right)=\stochasticmatrix_{Q}\left(t\right)\tensorprod\stochasticmatrix_{R}\left(t\right).\label{eq:TransitionMatrixParentSystemSubsystemsFactorizeSpacelikeSeparation}
\end{equation}
 Here $\stochasticmatrix_{Q}\left(t\right)$ is the unistochastic
transition matrix for the subsystem $Q$ corresponding to $\timeevop_{Q}\left(t\right)$
in the sense of the modulus-squaring relationship \eqref{eq:UnistochasticTransitionMatrixFromUnitaryEntries},
and $\stochasticmatrix_{R}\left(t\right)$ is similarly the unistochastic
transition matrix for the subsystem $R$ corresponding to $\timeevop_{R}\left(t\right)$.

It follows immediately from the tensor-factorization \eqref{eq:TransitionMatrixParentSystemSubsystemsFactorizeSpacelikeSeparation},
together with the definition \eqref{eq:TransitionMatrixAsConditionalProbabilities}
of the entries of a transition matrix as conditional probabilities,
that the overall system's directed conditional probabilities factorize
as 
\begin{equation}
p\left(\left(q_{t},r_{t}\right),t\given\left(q_{0},r_{0}\right),0\right)=p\left(q_{t},t\given q_{0},0\right)p\left(r_{t},t\given r_{0},0\right).\label{eq:ConditionalProbabilitiesParentSystemSubsystemsFactorizeSpacelikeSeparation}
\end{equation}
 Hence, marginalizing over $r_{t}$ leaves a conditional probability
for $Q$ that does not depend on $r_{0}$, precisely as in \eqref{eq:PairSubsystemsDirectedConditionalProbabilityNoCausalInfluence},
and a similar statement holds with $Q$ and $R$ switched. One can
therefore conclude that the principle of causal locality stated above
in \eqref{eq:PrincipleCausalLocality} is satisfied within this unistochastic
formulation of quantum theory.

By contrast, suppose that the two subsystems $Q$ and $R$ are \emph{not}
kept at spacelike separation during the physical process in question,
but locally interact at some intermediate time $t^{\prime}$ between
$0$ and $t$. Then, again following standard textbook arguments,
the overall system's unitary time-evolution operator $\timeevop_{QR}\left(t\right)$
will fail to tensor-factorize at $t^{\prime}$: 
\begin{equation}
\timeevop_{QR}\left(t^{\prime}\right)\ne\timeevop_{Q}\left(t^{\prime}\right)\tensorprod\timeevop_{R}\left(t^{\prime}\right).\label{eq:TimeEvOpParentSystemSubsystemsFailFactorizeFromInteraction}
\end{equation}
 Because the corresponding transition matrix $\stochasticmatrix_{QR}\left(t\right)$
encodes \emph{cumulative} statistical effects starting at the initial
time $0$, the transition matrix will \emph{continue} to fail to tensor-factorize
for all times $t\geq t^{\prime}$ (at least until the next division
event): 
\begin{equation}
\stochasticmatrix_{QR}\left(t\right)\ne\stochasticmatrix_{Q}\left(t\right)\tensorprod\stochasticmatrix_{R}\left(t\right)\quad\left[\textrm{for }t\geq t^{\prime}\right].\label{eq:TransitionMatrixParentSystemSubsystemsFailFactorizeFromInteraction}
\end{equation}

The breakdown \eqref{eq:TransitionMatrixParentSystemSubsystemsFailFactorizeFromInteraction}
in tensor-factorization for $t\geq t^{\prime}$ is precisely \emph{entanglement},
as manifested at the level of the underlying indivisible stochastic
process. The factorization \eqref{eq:ConditionalProbabilitiesParentSystemSubsystemsFactorizeSpacelikeSeparation}
therefore also breaks down, and so one can conclude that the two subsystems
$Q$ and $R$ exert causal influences on each other, stemming from
their local interaction at the time $t^{\prime}$.

Notice that this local interaction, despite being the \textquoteleft common
cause\textquoteright{} of the correlations between $Q$ and $R$,
is not the sort of \textquoteleft variable\textquoteright{} that can
be plugged into the unistochastic theory's microphysical conditional
probabilities. Reichenbach's principle of common causes \eqref{eq:ReichenbachCommonCausePrincipleFactorization}
therefore does not hold.

\section{Revisiting the Einstein-Podolsky-Rosen Argument\label{sec:Revisiting-the-Einstein-Podolsky-Rosen-Argument}}

\begin{figure}

\includegraphics[scale=0.6]{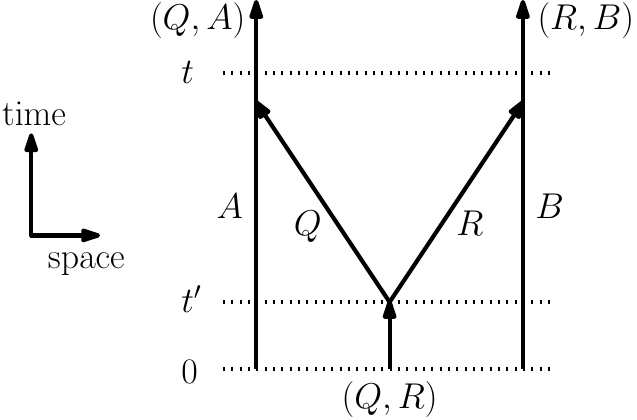}\caption{\label{fig:EPRExperimentSpacetimeDiagram}A spacetime diagram depicting
an idealized version of the EPR thought experiment, with the two subsystems
$Q$ and $R$ separating in space after they interact at the time
$t^{\prime}$, and then respectively joining up with the two observer-subsystems
$A$ (\textquoteleft Alice\textquoteright ) and $B$ (\textquoteleft Bob\textquoteright ).
The two observer-subsystems $A$ and $B$ are assumed to remain spacelike
separated throughout the experiment.}

\end{figure}

The stage is now set for revisiting the EPR argument. Referring to
Figure~\ref{fig:EPRExperimentSpacetimeDiagram}, suppose that an
observer $A$ (\textquoteleft Alice\textquoteright ) has local access
to the first subsystem $Q$, and that an observer $B$ (\textquoteleft Bob\textquoteright )
has local access to the second subsystem $R$, with no assumption
that $A$ and $B$ are in local contact with each other. Treating
$A$ and $B$ as ordinary (if complicated) subsystems of the overall
stochastic process, one now has a transition matrix of the form 
\begin{equation}
\stochasticmatrix_{QRAB}\left(t\right),\label{eq:OverallTransitionMatrixForEPR}
\end{equation}
 with individual entries consisting of directed conditional probabilities
of the form 
\begin{equation}
p\left(\left(q_{t},r_{t},a_{t},b_{t}\right),t\given\left(q_{0},r_{0},a_{0},b_{0}\right),0\right).\label{eq:OverallConditionalProbabilitiesForEPR}
\end{equation}
 (Note that $A$ and $B$ here do not denote random variables or observables,
but refer to subsystems.) 

The calculations ahead, which will be closely related to the no-communication
theorem~\citep{GhirardiRiminiWeber:1980agaastttqmmp,Jordan:1983qcdnts},
will show that the observer-subsystem $B$ does not exert a causal
influence on the observer-subsystem $A$. By symmetry, it will also
follow that $A$ does not exert a causal influence on $B$.

One begins by expressing the directed conditional probabilities \eqref{eq:OverallConditionalProbabilitiesForEPR}
in the usual Hilbert-space formalism as the following trace: 
\begin{equation}
\tr\left(\projector_{q_{t},r_{t},a_{t},b_{t}}\,\densitymatrix_{QRAB}\left(t\right)\right).\label{eq:OverallConditionalProbabilitiesForEPRAsHilbertSpaceTrace}
\end{equation}
 Here $\projector_{q_{t},r_{t},a_{t},b_{t}}$ is a rank-one projector
onto the state vector $\ket{q_{t},r_{t},a_{t},b_{t}}$, 
\begin{equation}
\projector_{q_{t},r_{t},a_{t},b_{t}}\defeq\ket{q_{t},r_{t},a_{t},b_{t}}\bra{q_{t},r_{t},a_{t},b_{t}},\label{eq:ProjectorForFinalStateForEPR}
\end{equation}
 and $\densitymatrix_{QRAB}\left(t\right)$ is the overall system's
density matrix at the time $t$, 
\begin{equation}
\densitymatrix_{QRAB}\left(t\right)\defeq\timeevop_{QRAB}\left(t\right)\densitymatrix_{QRAB}\left(0\right)\timeevop_{QRAB}^{\adj}\left(t\right),\label{eq:FinalDensityMatrixForEPR}
\end{equation}
 with $\densitymatrix_{QRAB}\left(0\right)$ the initial density matrix
at the time $0$, 
\begin{equation}
\densitymatrix_{QRAB}\left(0\right)\defeq\ket{q_{0},r_{0},a_{0},b_{0}}\bra{q_{0},r_{0},a_{0},b_{0}},\label{eq:InitialDensityMatrixForEPR}
\end{equation}
 and with $\timeevop_{QRAB}\left(t\right)$ the unitary time-evolution
operator for the overall system.

Suppose that the two subsystems $Q$ and $R$ locally interact only
at a time $t^{\prime}>0$. Then one can rewrite the formula \eqref{eq:FinalDensityMatrixForEPR}
for the overall system's density matrix at the later time $t\geq t^{\prime}$
as 
\begin{equation}
\densitymatrix_{QRAB}\left(t\right)\defeq\timeevop_{QRAB}\left(t\from t^{\prime}\right)\densitymatrix_{QRAB}\left(t^{\prime}\right)\timeevop_{QRAB}^{\adj}\left(t\from t^{\prime}\right).\label{eq:DensityMatrixForEPRStartingFromInteractionTime}
\end{equation}
 Here $\timeevop_{QRAB}\left(t\from t^{\prime}\right)$ is the relative
time-evolution operator for the time interval from $t^{\prime}$ to
$t$, defined as in \eqref{eq:DefRelativeTimeEvOp}, and $\densitymatrix_{QRAB}\left(t^{\prime}\right)$
is the overall system's density matrix at the interaction time $t^{\prime}$,
\begin{align}
\densitymatrix_{QRAB}\left(t^{\prime}\right) & \defeq\timeevop_{QRAB}\left(t^{\prime}\right)\densitymatrix_{QRAB}\left(0\right)\timeevop_{QRAB}\left(t^{\prime}\right).\nonumber \\
 & =\ket{\Psi_{QR},a_{0},b_{0}}\bra{\Psi_{QR},a_{0},b_{0}},\label{eq:DensityMatrixForEPRAtInteractionTime}
\end{align}
 with $\Psi_{QR}$ denoting the (now-entangled) wave function of the
subsystem pair $\left(Q,R\right)$.

By assumption, the relative time-evolution operator $\timeevop_{QRAB}\left(t\from t^{\prime}\right)$
from $t^{\prime}$ to $t$ encodes local interactions between the
two subsystems $Q$ and $A$, as well as local interactions between
the two subsystems $R$ and $B$, but no local interactions between
the subsystem pair $\left(Q,A\right)$ and the subsystem pair $\left(R,B\right)$.
Hence, the relative time-evolution operator tensor-factorizes as 
\begin{equation}
\timeevop_{QRAB}\left(t\from t^{\prime}\right)=\timeevop_{QA}\left(t\from t^{\prime}\right)\tensorprod\timeevop_{RB}\left(t\from t^{\prime}\right).\label{eq:RelativeTimeEvOpAfterInteractionForEPRFactorizes}
\end{equation}

It follows from a straightforward calculation that the reduced density
matrix for the subsystem pair $\left(Q,A\right)$ at the later time
$t\geq t^{\prime}$ is given by 
\begin{align}
 & \densitymatrix_{QA}\left(t\right)\defeq\tr_{RB}\left(\densitymatrix_{QRAB}\left(t\right)\right)\nonumber \\
 & =\tr_{RB}\bigg(\Bigparens{\timeevop_{QA}\left(t\from t^{\prime}\right)\tensorprod\timeevop_{RB}\left(t\from t^{\prime}\right)}\nonumber \\
 & \qquad\densitymatrix_{QRAB}\left(t^{\prime}\right)\Bigparens{\timeevop_{QA}^{\adj}\left(t\from t^{\prime}\right)\tensorprod\timeevop_{RB}^{\adj}\left(t\from t^{\prime}\right)}\bigg)\nonumber \\
 & =\tr_{R}\bigg(\Bigparens{\timeevop_{QA}\left(t\from t^{\prime}\right)\tensorprod\idmatrix_{R}}\nonumber \\
 & \qquad\Bigparens{\ket{\Psi_{QR},a_{0}}\bra{\Psi_{QR},a_{0}}}\Bigparens{\timeevop_{QA}^{\adj}\left(t\from t^{\prime}\right)\tensorprod\idmatrix_{R}}\bigg),\label{eq:ReducedDensityMatrixFirstPairSubsystemsForEPR}
\end{align}
 where $\idmatrix_{R}$ is the identity operator on the Hilbert space
of the subsystem $R$.  Notice that all the dependence on $b_{0}$
has disappeared. Thus, upon marginalizing over $q_{t}$, $r_{t}$,
and $b_{t}$, one finds 
\begin{align}
 & p\left(a_{t},t\given\left(q_{0},r_{0},a_{0},b_{0}\right),0\right)\nonumber \\
 & =\sum_{q_{t},r_{t},b_{t}}p\left(\left(q_{t},r_{t},a_{t},b_{t}\right),t\given\left(q_{0},r_{0},a_{0},b_{0}\right),0\right)\nonumber \\
 & =p\left(a_{t},t\given\left(q_{0},r_{0},a_{0}\right),0\right),\label{eq:ConditionalProbabilityFirstObserverForEPR}
\end{align}
 where 
\begin{align}
 & p\left(a_{t},t\given\left(q_{0},r_{0},a_{0}\right),0\right)\nonumber \\
 & \defeq\bra{a_{t}}\tr_{QR}\bigg(\Bigparens{\timeevop_{QA}\left(t\from t^{\prime}\right)\tensorprod\idmatrix_{R}}\nonumber \\
 & \qquad\Bigparens{\ket{\Psi_{QR},a_{0}}\bra{\Psi_{QR},a_{0}}}\Bigparens{\timeevop_{QA}^{\adj}\left(t\from t^{\prime}\right)\tensorprod\idmatrix_{R}}\bigg)\ket{a_{t}}.\label{eq:ConditionalProbabilityFirstObserverForEPRDetailed}
\end{align}
 One sees explicitly that there is no causal influence on the observer-subsystem
$A$ from the observer-subsystem $B$, in the sense of causal influences
used in this paper. The only causal influences on the observer-subsystem
$A$ are from the two subsystems $Q$ and $R$, which both intersect
the past light cone of $A$.

\section{Conclusion\label{sec:Conclusion}}

The past century has seen the appearance of many interpretations of
quantum theory, nearly all of which treat the wave function and the
Schr{\"o}dinger equation as the central entities of the theory, and differ
on whether to regard the wave function as a physical object. As a
purportedly physical object, the wave function would presumably be
understood to be some sort of field on a configuration space of very
high dimension, as Schr{\"o}dinger originally imagined in his early work
on what he called \textquoteleft undulatory mechanics\textquoteright ~\citep{Schrodinger:1926autotmoaam},
or as existing in an abstract Hilbert space of some very high dimension,
as might be more in keeping with Everett's \textquoteleft many worlds\textquoteright{}
interpretation~\citep{Everett:1957rsfqm,Everett:1973tuwf,DeWitt:1970qmr}.
Other approaches either augment the wave function with additional
(\textquoteleft hidden\textquoteright ) variables, like the pilot-wave
approach of de Broglie and Bohm~\citep{deBroglie:1930iswm,Bohm:1952siqtthvi,Bohm:1952siqtthvii},
or insist that the wave function is merely an instrumentalist tool
for encoding epistemic information about measurement settings and
results, as in some versions of the Copenhagen interpretation~\citep{Heisenberg:1958paptrims}.

None of these approaches provide a particularly hospitable domain
for talking about causation. They either rely inextricably on an interventionist
conception of causation, or they simply lack the kinds of microphysical
ingredients that merit being given causal meanings.

As explained above, the \emph{unistochastic formulation} of quantum
theory reviewed in this paper lies outside the wave-function paradigm,
and is based on treating every quantum system as a unistochastic process
in disguise~\citep{Barandes:2023tsqc,Barandes:2023tsqt}, an approach
that deflates a lot of the exotic talk about quantum phenomena. The
laws of this unistochastic process take the form not of differential
equations, but of directed conditional probabilities, which have a
long history of admitting an interpretation as encoding causal relationships.
From this perspective, quantum theory could be understood as a theory
of microphysical causation \emph{par excellence}.

By invoking this microphysical notion of causation, one can formulate
a more straightforward criterion \eqref{eq:PrincipleCausalLocality}
for causal locality than Bell's principle of local causality\textemdash in
either of its equivalent forms \eqref{eq:1975BellPrincipleOfLocalCausalityAsScreenOffCondition}
or \eqref{eq:1975BellPrincipleOfLocalCausalityAsFactorization}. As
this paper has shown, quantum theory, regarded as a theory of unistochastic
processes, satisfies this improved criterion, and is therefore arguably
a causally local theory. Remarkably, one therefore arrives at what
appears to be a causally local hidden-variables formulation of quantum
theory, despite many decades of skepticism that such a theory could
exist.

\section*{Acknowledgments}

The author would especially like to thank Isaac Friend, Wayne Myrvold,
Travis Norsen, John Norton, and Ward Struyve for helpful discussions.

\bibliographystyle{3_home_jacob_Documents_Work_My_Papers_Stochasti____Quantum_Theory__2023__custom-abbrvunsrturl}
\bibliography{2_home_jacob_Documents_Work_My_Papers_Bibliography_Global-Bibliography}

\end{document}